\documentclass[10pt,conference]{IEEEtran}
\IEEEoverridecommandlockouts
\usepackage{cite}
\usepackage{amsmath,amssymb,amsfonts}
\usepackage{algorithmic}
\usepackage{graphicx}
\usepackage{textcomp}
\usepackage{xcolor}
\usepackage{xcolor}  
\definecolor{mycitecolor}{RGB}{150, 30, 80}  
\usepackage[hyphens]{url}
\usepackage{makecell}
\usepackage{tabularx}
\usepackage{ragged2e}
\usepackage{booktabs}
\usepackage{multirow}
\usepackage{array}
\usepackage{adjustbox}
\usepackage{pifont}
\usepackage{enumitem}
\usepackage{subcaption}
\usepackage{graphbox}
\usepackage{stfloats}
\usepackage{caption}
\usepackage{float}
\usepackage[bookmarks=true,breaklinks=true,letterpaper=true,colorlinks,citecolor=mycitecolor,linkcolor=black,urlcolor=blue]{hyperref}
\usepackage[hyphenbreaks]{breakurl}
\usepackage{cleveref}
\usepackage{fancyhdr}
\usepackage{xspace}
\usepackage{marvosym}

\newcommand{\hpcayear}{2025}

\newcommand{\halfcheck}{\ding{52}\rotatebox[origin=c]{-9.2}{\kern-0.7em\ding{56}}}  
\newcommand{\hpcasubmissionnumber}{315}
\title{\huge{Gaze into the Pattern: Characterizing Spatial Patterns with Internal Temporal Correlations for Hardware Prefetching}}

\def\hpcacameraready{}

\newcommand\hpcaauthors{
    Zixiao Chen,
    Chentao Wu\textsuperscript{\Letter}\thanks{\Letter~Corresponding author.},
    Yunfei Gu,
    Ranhao Jia,
    Jie Li, and
    Minyi Guo
}

\newcommand\hpcaaffiliation{
    \textit{Dept. of Computer Science and Engineering}\\
    \textit{Shanghai Jiao Tong University}
}
\newcommand\hpcaemail{
    \texttt{\{chen\_zx, wuct\}@sjtu.edu.cn}
}

\author{
  \ifdefined\hpcacameraready
    \IEEEauthorblockN{\hpcaauthors{}}
      \IEEEauthorblockA{
        \hpcaaffiliation{} \\
        \hpcaemail{}
      }
  \else
    \IEEEauthorblockN{\normalsize{HPCA \hpcayear{} Submission
      \textbf{\#\hpcasubmissionnumber{}}} \\
      \IEEEauthorblockA{
        Confidential Draft \\
        Do NOT Distribute!!
      }
    }
  \fi 
}

\fancypagestyle{camerareadyfirstpage}{%
  \fancyhead{}
  
  \fancyhead[C]{
    \ifdefined\aeopen
    \parbox[][12mm][t]{13.5cm}{\hpcayear{} IEEE International Symposium on High-Performance Computer Architecture (HPCA)}    
    \else
      \ifdefined\aereviewed
      \parbox[][12mm][t]{13.5cm}{\hpcayear{} IEEE International Symposium on High-Performance Computer Architecture (HPCA)}
      \else
      \ifdefined\aereproduced
      \parbox[][12mm][t]{13.5cm}{\hpcayear{} IEEE International Symposium on High-Performance Computer Architecture (HPCA)}
      \else
      \parbox[][0mm][t]{13.5cm}{\hpcayear{} IEEE International Symposium on High-Performance Computer Architecture (HPCA)}
    \fi 
    \fi 
    \fi 
    \ifdefined\aeopen 
      \includegraphics[width=12mm,height=12mm]{ae-badges/open-research-objects.pdf}
    \fi 
    \ifdefined\aereviewed
      \includegraphics[width=12mm,height=12mm]{ae-badges/research-objects-reviewed.pdf}
    \fi 
    \ifdefined\aereproduced
      \includegraphics[width=12mm,height=12mm]{ae-badges/results-reproduced.pdf}
    \fi
  }
  \fancyfoot[C]{}
}
\begin{document}
\maketitle

\ifdefined\hpcacameraready 
  \thispagestyle{camerareadyfirstpage}
  \pagestyle{empty}
\else
  \thispagestyle{plain}
  \pagestyle{plain}
\fi

\newcommand{\hpcaheight}{0mm}
\ifdefined\eaopen
\renewcommand{\hpcaheight}{12mm}
\fi

\begin{abstract}

    Hardware prefetching is one of the most widely-used techniques for hiding long data access latency.
    To address the challenges faced by hardware prefetching, architects have proposed to detect and exploit the spatial locality at the granularity of spatial region.
    When a new region is activated, they try to find similar previously accessed regions for footprint prediction based on system-level environmental features such as the trigger instruction or data address.
    However, we find that such context-based prediction cannot capture the essential  characteristics of access patterns, leading to limited flexibility, practicality and suboptimal prefetching performance.

    In this paper, inspired by the temporal property of memory accessing, we note that the temporal correlation exhibited within the spatial footprint is a key feature of spatial patterns.
    To this end, we propose Gaze, a simple and efficient hardware spatial prefetcher that skillfully utilizes footprint-internal temporal correlations to efficiently characterize spatial patterns.
    Meanwhile, we observe a unique unresolved challenge in utilizing spatial footprints generated by spatial streaming, which exhibit extremely high access density.
    Therefore, we further enhance Gaze with a dedicated two-stage approach that mitigates the over-prefetching problem commonly encountered in conventional schemes.
    Our comprehensive and diverse set of experiments show that Gaze can effectively enhance the performance across a wider range of scenarios. Specifically, Gaze improves performance by 5.7\% and 5.4\% at single-core, 11.4\% and 8.8\% at eight-core, compared to most recent low-cost solutions PMP and vBerti.

\end{abstract}

\section{Introduction}
\label{sec:introduction}

Over the past few decades, the capacity of main memory has experienced exponential growth; however, the latency gap between the memory and CPU, known as the Memory Wall, has not seen significant improvement~\cite{computer_architecture_a_quantitative_approach, memory_wall_1,memory_wall_2}.
Furthermore, this performance disparity is becoming increasingly critical with the surge of memory-intensive applications such as big data~\cite{big_data} and deep learning~\cite{deep_learning}.

Hardware prefetching is a widely-used and extensively-studied approach that aims to bridge this gap~\cite{Primer,computer_architecture_a_quantitative_approach}.
By speculating and fetching data blocks into the cache proactively prior to explicit CPU demands,
data prefetching can hide the long data access latency and alleviate the pressure on the memory subsystem~\cite{redundant_prefetches_2}.
Hardware prefetchers try to extract memory access patterns from observed memory requests. When the learned pattern is likely to repeat, they can attempt to fetch possible future data~\cite{Primer}. Therefore, accurately characterizing program behavior profoundly impacts the performance of prefetching.

However, in practice, the diverse access behaviors and interferences such as out-of-order scheduling make it extremely challenging to achieve efficient prefetching while maintaining hardware simplicity.
To address these challenges, architects have proposed detecting and exploiting spatial locality~\cite{BOP,IPCP,SPP,SPP-PPF,Berti,Bingo,SMS,SMS_R,PMP,SPP_04,StealthPrefetching,BuMP,SFP,DSPatch,VLDB,Sandbox,Sangam,STeMS}. One effective approach is to utilize spatial patterns at the granularity of memory region (e.g., 4KB physical page)~\cite{Bingo,SMS,SMS_R,PMP,DSPatch,BuMP,SFP,STeMS,SPP_04,StealthPrefetching}. The spatial pattern of a region is represented as a bit vector of its spatial footprint~\cite{SPP_04}.
Spatial-pattern-based prefetchers learn spatial patterns by tracking several active memory regions and then use these patterns to predict the possible footprints of newly activated regions.
The effectiveness of spatial footprint prediction has been demonstrated in many prior studies.
It can effectively eliminate compulsory misses~\cite{Primer,PMP,Bingo,SPP_04,SFP} and is applicable in the presence of similar loops~\cite{PMP}, repetitive data layouts~\cite{SMS,Bingo,Primer,FootprintCache_2}, and data structure traversals~\cite{SMS,Bingo}. In addition, it can remain robust against out-of-order scheduling~\cite{DSPatch}.

To accurately identify similar regions for footprint prediction,
many researchers propose to characterize spatial patterns using system-level environmental features, such as the trigger instruction or data address.\footnote{Recent methods refer to the first access that touches a region as the trigger access. We use term `environmental features' to represent the features extracted from the trigger access, as they do not include internal information of its access pattern.} This is because there is a high probability of pattern recurrence if the system exhibits similar features when accessing a new region~\cite{SFP}.
For example, when using \texttt{PC+Address}, the recurrence indicates that the same instruction is used to touch the same address, suggesting a high likelihood of reproducing the same footprint~\cite{Bingo}.

\begin{figure}[t]
      \centering
      \includegraphics{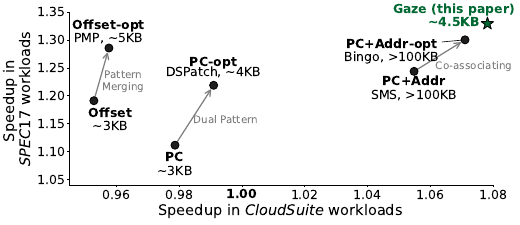} 
      \caption{Speedup achieved by different context-based characterization schemes and their hardware overheads. Suffix \textit{-opt} means an optimized version from recent literature.
      }
      \label{fig:fig1}
      \vspace{-10pt}
\end{figure}

Although such context-based characterization is widely used in microarchitectural techniques such as branch prediction~\cite{TAGE}, value prediction~\cite{ValuePrediction_1}, and cache replacement policies~\cite{CacheReplacementPolicies}, it is not as efficient as we expect for spatial pattern prediction. Figure~\ref{fig:fig1} plots the performance achieved by several characterization schemes for spatial-pattern-based prefetching. The detailed methodology is described in \S\ref{subsec:evaluation_methodology}.
The suffix `-opt' indicates an optimized version from recent literature. The x-axis shows the results of scale-out server workloads from \texttt{CloudSuite}~\cite{CloudSuite}, which contain a large amount of irregular accesses. The y-axis presents the results obtained from traditional \texttt{SPEC17}~\cite{SPEC17}. Prefetchers that combine both PC and data address for fine-grained characterization achieve excellent speedup but require substantial hardware costs. Conversely, characterizing at a coarse level, which is more hardware-friendly, results in numerous mispredictions in complex workloads.
Additionally, the optimized methods still fail to resolve the corresponding issues, seriously limiting the flexibility and practicality.

In this paper, we rethink how to intrinsically characterize spatial patterns. Given that access behaviors often exhibit temporal correlations within themselves~\cite{temporal_correlation}, the spatial patterns generated by them should also share similar properties. Inspired by this, we observe that when a spatial pattern recurs, its internal temporal correlation is likely to be reproduced in a similar manner. That is, the access order exhibited within the footprint plays a crucial role in characterizing spatial patterns.

In the light of this, we propose Gaze\footnote{The name Gaze is inspired by the idea of gazing into spatial patterns.}, a spatial prefetcher that innovatively utilizes this footprint-internal temporal correlation to efficiently characterize spatial patterns.
However, fully leveraging this temporal property is infeasible. Instead, we demonstrate that focusing on only several initial accesses can be highly efficient.
Based on this insight, Gaze chooses to utilize a small portion of internal temporal correlation to keep simplicity.
Furthermore, bit-vector-represented footprints in conventional spatial-pattern-based prefetchers do not capture any temporal information regarding the access order. Rather than introducing extra metadata, Gaze seamlessly incorporates temporal information into the experience searching process, thus enhancing both simplicity and compatibility with existing hardware designs.

Furthermore, we observe that footprints generated by spatial streaming typically exhibit extremely high access density. As it indicates to prefetch almost the entire region, a misuse of this pattern can lead to severe over-prefetching. However, being overly conservative may miss numerous prefetching opportunities. To address this unique challenge, we further optimize Gaze with a dedicated two-stage approach to dynamically adjust the prefetch aggressiveness.

We evaluate Gaze using various benchmarks that cover a variety of real-world applications.
Our evaluation encompasses various scenarios, including single-core, multi-core and bandwidth-constrained environments.
By extensive evaluation with seven state-of-the-art prefetchers we show that Gaze can effectively enhance the performance across a wider range of scenarios. Specifically, Gaze outperforms PMP and vBerti~\cite{Berti} by 5.7\% and 5.4\% on our entire single-core evaluation set respectively,
and by 11.4\% and 8.8\% on eight-core simulations. Additionally, the storage overhead of Gaze is almost the same with PMP, $31\times$ lower than Bingo~\cite{Bingo}. The area and energy overhead are much lower than that required by PMP and vBerti.

We make the following contributions in this paper:
\begin{itemize}
      \item
            We show that the weak link of spatial-pattern-based prefetching is its environmental-context-based pattern characterization, which seriously limits the flexibility and practicality.

      \item
            By leveraging the temporal correlation exhibited within the access footprint to characterize spatial patterns, the proposed Gaze can predict upcoming footprints more accurately. In addition, Gaze integrates this temporal information into the experience searching process, thus maintaining hardware simplicity.

      \item
            We further enhance Gaze for spatial streaming with a two-stage strategy to balance the demand for the entire region and the associated over-aggressiveness.

      \item
            We evaluate Gaze as well as seven recent prefetchers over 201 traces that cover a variety of real-world applications from \texttt{SPEC06}, \texttt{SPEC17}, \texttt{Ligra}, \texttt{PARSEC} and \texttt{CloudSuite} benchmark suites. Our evaluation shows that Gaze effectively enhances the performance in various scenarios.
\end{itemize}

\section{Related Works and Motivation}
\label{sec:background_and_motivation}

    \subsection{Recent Spatial-Pattern-based Mechanisms}
    \label{subsec:recent_prefetching_mechanisms}

    Spatial-pattern-based prefetching learns the spatial pattern of a certain region through directly recording its footprint and tries to find other similar regions to deploy the learned experience for prefetching~\cite{SPP_04,SFP}. The potential of it has been demonstrated by many prior works~\cite{SFP,StealthPrefetching,SPP_04,DSPatch,SMS,SMS_R,Bingo,PMP,BuMP,STeMS}.

    Spatial Footprint Predictor (SFP)~\cite{SFP} first exemplifies the common structure and logic of spatial-pattern-based prefetching. SFP proposes to use PC as well as some bits from the data address to characterize patterns, which improves the prefetch accuracy and enables the prefetcher to eliminate compulsory misses. Spatial Pattern Prediction~\cite{SPP_04} improves SFP with larger region size and cache line size.

    \textbf{Fine-grained spatial prefetching} typically characterizes patterns with several environmental contexts such as the activating instruction \textit{and} data address.
    Spatial Memory Streaming (SMS)~\cite{SMS} proposes to use \texttt{PC+Offset\footnote{Offset means the distance of the block address from the beginning of a region.}}. SMS also uses a filter table (FT) to filter out one-bit spatial footprints. Bulk Memory Access Prediction and Streaming~\cite{BuMP} further reduces the energy consumption of SMS.

    Bingo~\cite{Bingo} is motivated by the fact that the shorter event (\texttt{PC+Offset}) is carried in the longer one (\texttt{PC+Address}). Inspired by the TAGE~\cite{TAGE} branch predictor, Bingo first tries to use the longer trigger event to find the exact match. If it is not found, then the shorter event is used to find the approximate match.
    Therefore, compared to SMS, Bingo achieves similar accuracy (sustained by exact matches) with more misses being covered (approximate matches).

    Combining both trigger instruction and data address accurately captures the contextual features where the pattern begins.
    Therefore, predictions for matched regions typically achieve high accuracy.
    However, similar patterns are often associated with different events due to minor contextual differences, causing severe data redundancy~\cite{Bingo, PMP}.
    Thus, to achieve considerable coverage, substantial storage overhead and lengthy learning period are required~\cite{PMP,DSPatch}, making the prefetcher impractical. Both SMS and Bingo require more than 100KB of storage when they achieve their optimal performance~\cite{SMS,Bingo}.

    \textbf{Coarse-grained spatial prefetching.} To this point, several recent studies attempt to do it in simpler ways. They opt for coarse-grained events that are more hardware-friendly, and employ various methods to mitigate the resulting low accuracy.
    Coarse-grained methods characterize patterns with only a few bits from the trigger instruction \textit{or} data address.

    Dual Spatial Pattern Prefetcher (DSPatch)~\cite{DSPatch} characterizes spatial patterns at the granularity of instructions (i.e., \texttt{PC} is used). At the same time, it maintains two up-to-date patterns for each PC and can further improve accuracy/coverage by prefetching data blocks that appear in both/either pattern.

    Pattern Merging Prefetcher (PMP)~\cite{PMP} further coarsens the characterization by using only the \texttt{Offset}, ensuring that a match can almost always be found after working for a short period. This almost minimizes the chances of losing prefetch opportunities due to the execution of new instructions or access to unseen addresses.
    In addition, for each offset, PMP merges the 32 most recent patterns, meaning that their common features are preserved and utilized.

    \subsection{Existing Limitations and Motivation}

    Conventional context-based methods exhibit significant limitations in capturing key access characteristics. As shown in Figure~\ref{fig:fig1}, low-cost solutions are challenging to apply to complex workloads, while introducing more environmental contexts can lead to prohibitive hardware costs. Although numerous efforts, such as bandwidth utilization-aware aggressiveness adjustment~\cite{DSPatch}, pattern merging~\cite{PMP}, and long-short events co-associating~\cite{Bingo}, have been made to mitigate this inefficiency, their shortcomings remain unresolved, seriously limiting the flexibility and practicality of spatial-pattern-based prefetching.
    This motivates a deeper investigation into the patterns themselves to develop an efficient characterization scheme.

    Memory access often exhibits temporal locality, meaning that accessed data are likely to be reused in the near future~\cite{Primer}.
    Temporal correlation, as an analog to temporal locality, refers to the phenomenon where a group of addresses tends to be reused \textit{in the same order}~\cite{temporal_correlation}. It is widely used in identifying irregular patterns such as pointer chasing and graph processing~\cite{STeMS,Markov,STMS,GHB,ISB,Domino,Triage,MISB}.
    Since spatial patterns are generated by memory access, they should also exhibit similar temporal properties within themselves.
    This suggests that pattern recurrence may correlate with a consistent order of internal accesses.

    \begin{figure}[t]
        \centering
        \includegraphics{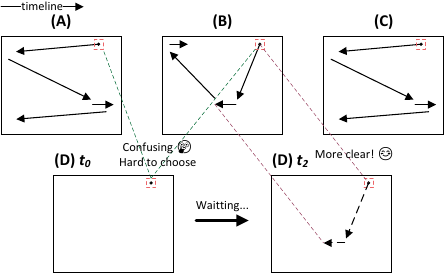} 
        \caption{The detailed reference footprints of several spatial regions. These regions are accessed close in time. The trigger accesses are cycled in red. The footprint of region \textit{B} is partially given.
        }
        \label{fig:fig2}
        \vspace{-10pt}
    \end{figure}

    The upper part of Figure~\ref{fig:fig2} provides an example obtained from \texttt{fotonik3d\_s}. It plots detailed footprints of three spatial regions. The spatial footprints as well as their internal temporal correlations exhibited in region \textit{A} and region \textit{C} are the same. That is, their referenced blocks are aligned and are accessed in the same order.
    In contrast, region \textit{A} and region \textit{B} show significant differences in terms of both spatial footprints and temporal access orders.

    Since footprint-internal temporal correlations are access-intrinsic, utilizing them for fine-grained characterization could be a feasible direction.
    To illustrate this, consider the scenario depicted in Figure~\ref{fig:fig2}.
    At time $t_0$ (see the lower part), regions \textit{A}, \textit{B} and \textit{C} have recently been deactivated, meaning their patterns have been learned by the prefetcher. Meanwhile, region \textit{D} is just activated at this point. The trigger accesses of these four regions, which are highlighted in red, are aligned. Thus, they are similar by definition. However, the patterns of region \textit{A}, \textit{B} and \textit{C} have met a conflict.
    This poses a critical challenge when choosing a suitable prefetch pattern for region \textit{D}. 
    Relying solely on the trigger offset is insufficient, and introducing additional context information like PC or full address incurs unacceptable metadata overhead.
    Our approach takes a different tack.
    Rather than focusing on trigger event refinement, we incorporate the information from footprint-internal accesses. 
    As shown in Figure~\ref{fig:fig2}, after just two subsequent accesses (at time $t_2$), the similarity between region \textit{B} and region \textit{D} becomes apparent. With this newfound insight, we can make a high-confidence prediction at time $t_2$. This enables fine-grained pattern characterization without the substantial increase in metadata overhead associated with combining various program contexts.

    However, this idea is not readily applicable to spatial-pattern-based prefetching, primarily due to \ding{202} its incompatibility with conventional hardware designs and \ding{203} the additional complexity it introduces.
    The first challenge arises because conventional methods make predictions based on trigger accesses, but to obtain temporal correlations, more observations are required.
    The second challenge is that storing and utilizing temporal information is expensive.
    Therefore, to maintain simplicity and efficiency, we need a careful design, or the effectiveness may be inferior to previous fine-grained approaches.

\section{Gaze Spatial Prefetcher}
\label{sec:gaze_spatial_prefetcher}

We propose Gaze, which to our best knowledge is the first spatial prefetcher that utilizes \textbf{footprint-internal temporal correlations} to efficiently characterize spatial patterns.
To balance hardware complexity and performance, Gaze leverages information from region's first two accesses (\S\ref{subsec:Pattern Characterization}). This simplification allows Gaze to seamlessly integrate temporal feature extraction into existing hardware designs without extra metadata storage.
At the same time, Gaze introduces a dedicated two-stage approach to mitigate the over-prefetching problem that often occurs in conventional schemes when they utilize high-dense footprints generated from spatial streaming (\S\ref{subsec:Enhancement towards Spatial Streaming}).

Gaze is designed and primarily evaluated as an L1D prefetcher. Nevertheless, it can also be placed at L2C, working in conjunction with the existing commercial L1D prefetcher IP-stride.

    \subsection{Design Overview}

    \begin{figure*}[htb]
        \centering
        \includegraphics{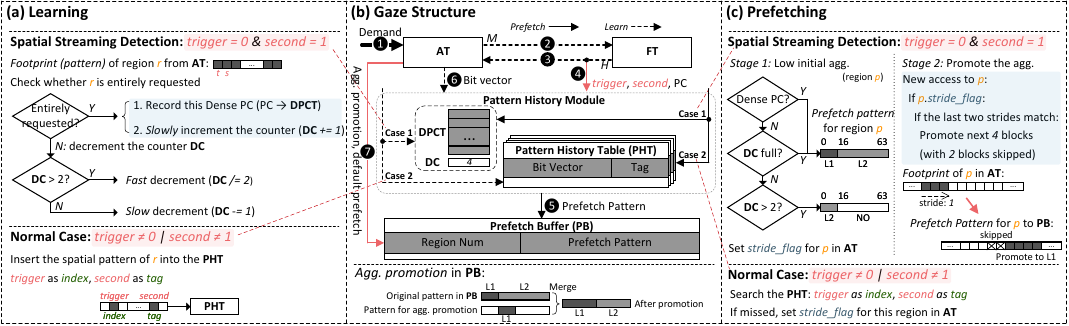} 
        \caption{Design overview of Gaze (b) and its detailed learning (a) and prefetching (c) process. For simplicity, we use \textit{trigger} and \textit{second} to refer to the trigger offset and the second offset, omitting the term \textit{offset}. \textit{H} and \textit{M} mean a hit and a miss in the corresponding structure, respectively.}
        \label{fig:Gaze}
        \vspace{-10pt}
    \end{figure*}

    Figure~\ref{fig:Gaze} illustrates the design of Gaze. The middle sub-figure (Figure~\ref{fig:Gaze}b) shows the design overview, while the left sub-figure (Figure~\ref{fig:Gaze}a) and the right sub-figure (Figure~\ref{fig:Gaze}c) illustrate the detailed learning and prefetching procedure, respectively.
    The key innovations, including the use of the temporal information from the first two accesses (the detail is discussed in \S\ref{subsec:Pattern Characterization}) and the path of dedicated aggressiveness control (\S\ref{subsec:Enhancement towards Spatial Streaming}), are highlighted in Figure~\ref{fig:Gaze}b.

    As shown in Figure~\ref{fig:Gaze}b, Gaze is trained on cache loads.
    Similar to conventional spatial-pattern-based prefetchers{~\cite{Bingo, PMP, SMS}}, Gaze consists of three main components: Filter Table (FT), Accumulation Table (AT), and Pattern History Module (PHM). FT is used to filter out one-bit spatial patterns, indicating that only one block in the region is demanded during tracking. AT is used to track all active regions. PHM is used to learn access pattern, and to issue prefetches based on the learned experience. Any newly activated region is first recorded in FT and will be tracked by AT only after the arrival of the second different load access. Gaze also employs a Prefetch Buffer (PB) to efficiently store the prefetch addresses, as a single bit vector often contains multiple requests with the same starting address, i.e., the region number. In addition, PB also helps to smooth the issuance of prefetches.

    \textbf{Access Flow.} Upon the arrival of a {load}, Gaze initially checks AT to see if the accessed region is under tracking (\ding{202}), and if so, Gaze updates the corresponding footprint.
    Otherwise, Gaze looks up the FT (\ding{203}). If the region is found, which means that it has been activated once before, Gaze then starts tracking it in AT (\ding{204}).
    Unlike previous proposals, Gaze will also send the trigger offset, the second offset as well as the trigger PC to PHM for prefetching (\ding{205}). This is because Gaze chooses to incorporate the temporal correlation between the first two accesses, which is available at this point. If more accesses are required, this step would be delayed.
    Conventional methods{~\cite{Bingo, PMP, SMS, DSPatch}} awaken the prefetching process through the trigger access rather than the second access.
    After receiving these, PHM decides whether to trigger prefetching for this region. If so, the corresponding prefetch pattern will be sent to PB, ready for prefetching (\ding{206}).  For each region that is under tracking, once it is deactivated (e.g., one of its cached blocks is evicted out of the cache, 
    or its associated tracking entry is evicted from the AT by the LRU policy due to inactivity), the accumulation ends and the bit vector is sent to PHM (\ding{207}) for pattern learning.

    Each entry in the AT is used to track an active region, where we introduce two additional 6-bit fields to store the offsets of the region's last two accesses, which, together with the new access, will collectively calculate the last two strides. AT decides whether to perform region-based stride prefetching (used as the backup prefetching) and aggressiveness promotion based on the two strides (\ding{208}).

    \subsection{Pattern Characterization}
    \label{subsec:Pattern Characterization}

    To identify footprint-internal features, Gaze has to wait for several initial accesses.
    Thus, the most critical design choice is the number of accesses used for temporal correlation extraction. 
    This requires Gaze to hold and observe all recently activated regions, differentiating their spatial patterns based on their recently accessed blocks in the presence of strong interference such as out-of-order scheduling. Besides, this also means the loss of prefetch opportunities. Therefore, waiting for too many accesses incurs negative effect.

\begin{figure}[htb]
        \centering
        \includegraphics{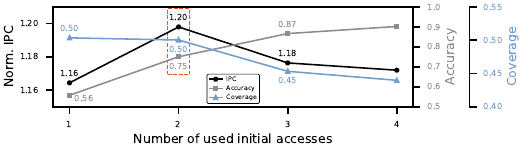} 
        \caption{Effect of extending the number of aligned initial accesses that required for a match. The region size is set to 4KB.}
        \label{fig:fig4}
    \end{figure}

    To strike a good performance-cost balance, we first explore the relationship between the amount of temporal information used and the resulting efficiency in terms of speedup, accuracy, and coverage. 
    We say two regions are similar when their initial accesses are aligned in both spatial footprint and temporal order. Figure~\ref{fig:fig4} illustrates the effect of extending the number of aligned initial accesses required from one to up to four. The results are the averages across the entire evaluation set. We sequentially concatenate the offsets of these accesses to form the index event. When only the trigger offset is used, the size of the pattern history table (PHT) is 64, since there are 64 distinct offsets within a 4KB region {and no instruction information is incorporated}. However, when considering more accesses, a full PHT becomes impractical and inefficient. Therefore, for simplicity, we use 256-entry fully-associative history tables in these cases, which consume less than 3KB.\footnote{The performance is saturated after the size is larger than 256.}

    As illustrated in Figure~\ref{fig:fig4}, if we require that the first four demanded blocks must be aligned both spatially and temporally, the prefetch accuracy increases from 56\% to 90\%, indicating the successful characterization of access behaviors.
    However, prefetch opportunities are substantially lost, resulting in a decrease in IPC and coverage. An ideal balance is achieved when we choose to only use the first two accesses, which provides a 3\% improvement in IPC and a 35\% improvement in accuracy, with only negligible reduction in coverage. When the number of used offsets increases again, a marginal gain in accuracy comes at a cost of a significant degradation in both IPC and coverage.

    Building on this insight, we decide to leverage the temporal correlation between the first two accesses to characterize spatial patterns. This choice offers three major advantages.
    Firstly, this approach significantly reduces storage overhead compared to conventional fine-grained methods that rely on diverse program contexts.
    Secondly, it balances performance, accuracy, and coverage, ensuring timely prefetches and minimizing interference from out-of-order scheduling. 
    Thirdly, it can be seamlessly integrated into recent hardware designs.
    Gaze utilizes existing logic to capture the first two offsets and store patterns in a way that does not cost extra metadata overhead, maintaining simplicity and compatibility.
    Specifically, recent methods employ FT to invalidate the regions that are accessed only once, which allows us to obtain these two offsets when FT sends a valid region to AT (\ding{205} in Figure~\ref{fig:Gaze}b). Additionally, we use the first offset as the index and the second offset as the tag to store patterns, with their order inherently verified during the table lookup process.

    In addition, to maintain its high accuracy, Gaze employs a \textit{strict matching mechanism}, which prevents awakening prefetching when only one access matches while the other does not. This differs slightly from the mechanisms employed in Bingo~\cite{Bingo}, TAGE~\cite{TAGE}, and Domino~\cite{Domino}, which enable prediction even in cases of partial matches. We restrict this ability primarily because the temporal correlation between the first two accesses is a key characteristic.
    However, we may miss prefetching opportunities when the candidate region fails to find a match, even though its upcoming pattern may be easy to follow. To compensate for this, we enhance the tracking structure with a {backup} prefetcher. The detail will be further discussed together with our dedicated two-stage aggressiveness control in \S\ref{subsec:Enhancement towards Spatial Streaming}.

    \subsection{Enhancement towards Spatial Streaming}
    \label{subsec:Enhancement towards Spatial Streaming}

\begin{figure}[tb]
        \centering
        \includegraphics{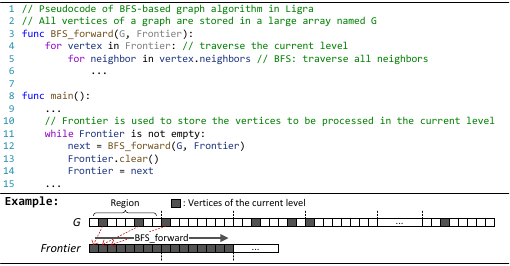} 
        \caption{Pseudocode of BFS-based graph processing and an example of its illustration.}
        \label{fig:fig5}
        \vspace{-10pt}
    \end{figure}

    Spatial streaming refers to the phenomenon of prolonged spatially-strided accessing~\cite{Primer}.
    Despite the extensive research on spatial streaming over the past decades~\cite{stream1, stream2, stream3, stream4, stream5, stream6}, we still observe a unique unresolved challenge in utilizing its spatial footprints.
    Specifically, we find that footprints generated by spatial streaming typically exhibit extremely high access density, especially when the stride is one, implying that when used for prefetching, speculative prefetches will be issued for almost all blocks in the candidate region.
    Clearly, spatial streaming could greatly benefit from this high aggressiveness. However, in practice, various access patterns are often exhibited in an interleaved manner, and if such highly dense footprints are misused, the resulting massive and dense useless prefetches could cause significant cache pollution and off-chip bandwidth bottlenecks.
    For example, Figure~\ref{fig:fig5} provides the pseudocode of BFS-based graph processing algorithm in \texttt{Ligra}. The algorithm allocates a temporary space called \texttt{Frontier} to store sparsely distributed vertices to be processed in the current level. Thus, traversing \texttt{Frontier} (in function \texttt{BFS\_forward}) yields access patterns that include both irregular accessing and spatial streaming.

    To overcome this challenge, like many previous approaches{~\cite{stream2,stream3,stream4}}, we use a PC-based mechanism. Specifically, we deploy a Dense PC Table (DPCT) to record recent dense PCs, and a Dense Counter (DC) to track the frequency of recent dense footprints. We perform a double-checking using these two structures, and a two-stage approach to slowly increment the prefetch aggressiveness.
    
    \textbf{Learning Phase.} Spatial streaming detection is illustrated in the upper part of Figure~\ref{fig:Gaze}a. When we finish tracking a region whose initial two accessed blocks are the block 0 and block 1 (i.e., \textit{r} in the figure), we check whether it is highly dense, i.e., whether all of its blocks have been accessed. If so, we record its trigger instruction into DPCT and increment the DC. Otherwise, we decrement DC, where a large value prompts a fast decrement.
    
    \textbf{Prefetching Phase.} We employ a two-stage approach to adjust the prefetching aggressiveness when trying to apply this dense pattern to the candidate region.
    In the first stage, for the candidate region, we detect the confidence level of exhibiting spatial streaming, and assign an initial prefetch aggressiveness accordingly. In the second stage, we dynamically adjust the prefetch aggressiveness based on subsequent accesses, gradually increasing it as needed.
    
    \textit{Stage 1.} As shown in the upper part of Figure~\ref{fig:Gaze}c, when issuing prefetches for the region \textit{p} that is likely to be fully demanded by spatial streaming, if the trigger instruction has been recently labelled as a dense PC or DC is saturated, we assign a moderate level of aggressiveness: prefetching the first 16 blocks into the L1D and the remaining blocks into the L2C. We empirically set the number of initial blocks with higher aggressiveness to 16 (which is one-quarter of the region) to avoid losing prefetch opportunities while not being overly aggressive. Otherwise, if DC is half-saturated, indicating a lower probability that \textit{p} will be a dense region, we prefetch only the first 16 blocks into the L2C. If neither the trigger instruction has been labelled nor the DC is big enough, we refrain from prefetching. Meanwhile, when inserting this region into AT, we set the \texttt{stride\_flag} mark. This enables AT to prefetch the remaining blocks if the region exhibits streaming behavior later.
    
    \textit{Stage 2.} AT will track all marked regions. For each one, we compute the two strides of the last three accesses. If the two strides both equal to one, we promote several subsequent blocks into the L1D (see the upper part of Figure~\ref{fig:Gaze}c). The lower part of Figure~\ref{fig:Gaze}b shows how the PB merges the promotion pattern and the original prefetch pattern.
    
    This stride-based approach can also serve as a {backup} predictor for cases where our strict matching mechanism (see \S\ref{subsec:Pattern Characterization}) fails to predict. As illustrated in the lower part of Figure~\ref{fig:Gaze}c, if no match is found, the \texttt{stride\_flag} of this region is also set during tracking. Generally, for marked regions, once the last two strides match, the region-based stride prefetching is activated.
    
    The differences between our region-based stride prefetching and other stride-based mechanisms are discussed below. Firstly, it serves a dual purpose: promoting aggressiveness and capturing potentially missed opportunities (\ding{208} in Figure~\ref{fig:Gaze}b). Secondly, it leverages the existing structure, e.g., the AT, and operates in a region rather than a global or per-PC view, avoiding the need for new tracking structures. 
    So, to maintain simplicity, complex scenarios such as multiple strides, are not handled.

    \subsection{Pattern History Module}
    \label{subsec:Pattern History Module}
    
    As outlined in Figure~\ref{fig:Gaze}b, in addition to the DPCT and DC, which are responsible for handling large spatial streaming mentioned in \S\ref{subsec:Enhancement towards Spatial Streaming} (denoted as case 1 for brevity), PHM also includes a Pattern History Table (PHT) used in typical scenarios (denoted as case 2). As previously mentioned, if the first two accesses to a region are spatially adjacent and occur at the head of the region, the DPCT and DC will learn this pattern. Otherwise, the PHT will be used to learn the pattern. Similarly, as for the footprint prediction of the candidate region, PHM invokes the corresponding structure based on its first two accessed blocks.
    Additionally, PHT prefetches all blocks into the L1D.
    
    \subsection{Hardware Overhead}
    \label{subsec:design_storage_requirements}

    \textbf{Storage overhead.} Gaze sets the region size to 4KB (the same as a typical physical page). Table~\ref{tab:tab1} presents the breakdown of each component and the total storage overhead. Both FT and AT have 64 entries, enabling the simultaneous tracking of 64 pages. PB uses 32 entries to store prefetch patterns for up to 32 pages, with four states for each offset: No Prefetch, Prefetch to L1D, L2C, and LLC (not used). PHT uses 4-way set-associativity. DPCT contains 8 entry. We omit the cost of DC since it only takes 3 bits.
    Lastly, the total hardware overhead of Gaze is 4.46KB ($31\times$ lower than Bingo~\cite{Bingo}, 0.54KB less than PMP~\cite{PMP}, 1.91KB more than Berti~\cite{Berti}).
    \begin{table}[htb]\renewcommand{\arraystretch}{1.5}
        \centering
        \caption{Detailed storage requirements}
        \label{tab:tab1}
        {
            \begin{tabular}{|p{0.45in}<{\centering}|p{2.05in}|p{0.4in}<{\centering}|}
                \hline
                \textbf{Structure}                   &
                \centering\textbf{Description}       &
                \textbf{Storage}                                         \\ \hline\hline
    
                FT                                   &
                \makecell[l]{\specialrule{0em}{0.5pt}{0.5pt}
                $8$-way; $64$ entries, each containing:                  \\
                $\bullet$ Region Tag ($36$b) \& LRU ($3$b)               \\
                $\bullet$ Hashed PC ($12$b) \& Trigger offset ($6$b)     \\
                    \specialrule{0em}{0.5pt}{0.5pt}
                }                                    &
                456B                                                     \\ \hline
    
                AT                                   &
                \makecell[l]{\specialrule{0em}{0.5pt}{0.5pt}
                $8$-way; $64$ entries, each containing:                  \\
                $\bullet$ Region Tag ($36$b) \& LRU ($3$b)               \\
                $\bullet$ Hashed PC ($12$b) \& stride\_flag ($1$b)       \\
                $\bullet$ Trigger \& Second offset ($2\times6$b)         \\
                $\bullet$ Last \& Penultimate offset ($2\times6$b)       \\
                $\bullet$ Bit vector ($64$b)                             \\
                    \specialrule{0em}{0.5pt}{0.5pt}
                }                                    &
                1128B                                                    \\ \hline
    
                PHT                                  &
                \makecell[l]{\specialrule{0em}{0.5pt}{0.5pt}
                $4$-way; $256$ entries, each containing:                 \\
                $\bullet$ Tag ($6$b) \& LRU ($2$b) \& Bit vector ($64$b) \\
                    \specialrule{0em}{0.5pt}{0.5pt}
                }                                    &
                2304B                                                    \\ \hline
    
                DPCT                                 &
                \makecell[l]{\specialrule{0em}{0.5pt}{0.5pt}
                Fully-associative, $8$ entries, each containing:         \\
                $\bullet$ Hashed PC ($12$b) \& LRU ($3$b)                \\
                    \specialrule{0em}{0.5pt}{0.5pt}
                }                                    &
                15B                                                      \\ \hline
    
                PB                                   &
                \makecell[l]{\specialrule{0em}{0.5pt}{0.5pt}
                $8$-way; $32$ entries, each containing:                  \\
                $\bullet$ Region Tag ($36$b) \& LRU ($3$b)               \\
                $\bullet$ Prefetch Pattern ($64\times 2$b)               \\
                    \specialrule{0em}{0.5pt}{0.5pt}
                }                                    &
                668B                                                     \\ \hline \hline
                \multicolumn{2}{|c|}{\textbf{Total}} & \textbf{4.46KB}   \\ \hline
            \end{tabular}
            \vspace{-10pt}
        }
    \end{table}
    
    \textbf{Area and energy overhead.}
    We estimate the area and access energy consumed by the pattern history module of Gaze (PHT and DPCT) and PMP (OPT and PPT) using CACTI~\cite{cacti1,cacti2} with its 22nm configuration. Gaze consumes about 29\% of the area compared to PMP (0.0034\thinspace  $mm^2$ vs. 0.0117\thinspace  $mm^2$). This is because each line in Gaze only costs 64b for storing a bit vector, where PMP costs 320b (160b) for storing a counter vector (coarse counter vector). Also, the read and write access to pattern history module of Gaze consumes less than 46\% of the energy required by PMP. Each table of Gaze can be accessed within a single CPU cycle.
    
    The area and energy consumption of Gaze are much lower than that of Berti, as Berti extends each L1D line by 12 bits to accommodate fetch latency. This extension incurs an over $10\times$ increase in both area and access energy compared to PHM in Gaze.

\section{Evaluation}
\label{sec:methodology}

    \subsection{Methodology}
    \label{subsec:evaluation_methodology}

    We evaluate Gaze along with other recent proposals using the ChampSim\footnote{Commit b4bfdb8. March 17, 2023.}~\cite{ChampSim}, a trace-driven cycle-accurate simulator capable of modeling various types of modern out-of-order processors.
    ChampSim is an effective framework to study microarchitectural techniques.
    This framework was used for the 1st instruction prefetching championship (IPC)~\cite{IPC1}, the 2nd and the 3rd data prefetching championships (DPC-2 and DPC-3)~\cite{DPC2, DPC3}, the 2nd cache replacement championship (CRC-2)~\cite{CRC2}, and ML-based data prefetching competition~\cite{ML-DPC}. Table~\ref{tab:tab2} presents the system configuration of our simulation.

    \begin{table}[htbp]\renewcommand{\arraystretch}{1.5}
        \centering
        \caption{Configuration of the simulator}
        \label{tab:tab2}
        {
            \begin{tabularx}{3.2in}{m{0.35in} m{2.6in}}
                \hline
                \textbf{Core} & 1-8 cores, 4GHz, 4-wide OoO, 128/72-entry LQ/SQ, 352-entry ROB, hashed perceptron branch predictor \\
                \hline
                \textbf{TLBs} & ITLB/DTLB: 64-entry, {1 cycle}; STLB: 1536-entry, {8 cycles}                                       \\
                \hline
                \textbf{L1I}  & 32KB, 64B line, 8 way, 4 cycles, 8 MSHRs                                                           \\
                \hline
                \textbf{L1D}  & 48KB, 64B line, 12 way, 5 cycles, 16 MSHRs                                                         \\
                \hline
                \textbf{L2C}  & 512KB, 64B line, 8 way, 10 cycles, 32 MSHRs                                                        \\
                \hline
                \textbf{LLC}  & 2MB/core, 64B line, 16 way, 20 cycles, 64 MSHRs                                                    \\
                \hline
                \textbf{DRAM} & \textbf{1C}: Single channel, 1 rank/channel; \textbf{2C}: Dual channel, 1 rank/channel;
                \textbf{4C}: Dual channel, 2 ranks/channel;
                \textbf{8C}: Quad channel, 2 ranks/channel;\newline
                DDR4~\cite{ddr}, 8 banks/rank, 3200 MTPS, 64b data-bus, 2KB row buffer/bank,
                tRP = tRCD = tCAS = 12.5ns                                                                               \\
                \hline
            \end{tabularx}
        }
    \end{table}

    \subsubsection{Workloads}
    We use memory-intensive workloads that cover a variety of real-world applications for evaluation. Our evaluation traces are captured from \texttt{SPEC} \texttt{CPU2006}~\cite{SPEC06}, \texttt{SPEC CPU2017}~\cite{SPEC17}, \texttt{PARSEC 2.1}~\cite{PARSEC}, \texttt{Ligra}~\cite{Ligra} and \texttt{CloudSuite}~\cite{CloudSuite} benchmark suites. For \texttt{SPEC CPU2006} and \texttt{SPEC CPU2017} workloads, we use the instruction traces provided by DPC-2 and DPC-3. For \texttt{PARSEC 2.1} and \texttt{Ligra} workloads, we use the traces open-sourced by Pythia~\cite{Pythia} and Hermes~\cite{Hermes}. For \texttt{CloudSuite} workloads, we use the traces from the CRC-2. We only evaluate those traces that are memory-insensive among these workloads, i.e., traces that have at least one miss per kilo instructions (MPKI) at the LLC when without prefetching. Table~\ref{tab:tab3} lists the number of traces for each suite, and some example workloads. In total, we evaluate 201 traces in our experiment.

    \begin{table}[tb]\renewcommand{\arraystretch}{1.2}
        \centering
        \caption{Evaluated workloads}
        \label{tab:tab3}
        {
            \begin{tabularx}{0.45\textwidth}{m{0.075\textwidth} c X}
                \toprule
                \textbf{Suite} & \textbf{\#Traces} & \textbf{Example Workloads}                \\
                \midrule
                SPEC06         & 39                & gcc, mcf, milc, soplex, sphinx3, leslie3d \\
                SPEC17         & 39                & gcc, mcf, bwaves, omnetpp, xalancbmk      \\
                Ligra          & 67                & pagerank, Bellman-ford, BFS, MIS          \\
                PARSEC         & 4                 & facesim, streamcluster                    \\
                Cloud          & 52                & nutch, cassandra, streaming               \\
                \bottomrule
            \end{tabularx}
        }
        \vspace{-10pt}
    \end{table}

    We use the first {200M} instructions to warmup, and the next 200M instructions to simulate per core. If a trace reaches its end before the simulator has executed enough instructions, it is replayed from the start. For multi-core simulations, any core that finishes early replays its trace until all cores have simulated enough instructions. We simulate both homogeneous and heterogeneous mixes in multi-core simulations. A homogeneous mix is composed of $n$ copies of a trace and a heterogeneous mix consists of $n$ randomly-selected traces, where $n$ is the number of cores. For $n$-core simulation, 201 homogeneous and $\lfloor\frac{201}{n}\rfloor$ heterogeneous mixes are evaluated. Each core runs one trace in the mix.

    \subsubsection{Prefetchers}

    In addition to Gaze, we evaluate seven state-of-the-art spatial prefetchers: SMS~\cite{SMS}, Bingo~\cite{Bingo}, DSPatch~\cite{DSPatch}, PMP~\cite{PMP}, IPCP~\cite{IPCP}, SPP-PPF~\cite{SPP, SPP-PPF} as well as Berti~\cite{Berti}.
    The first four prefetchers are based on spaital pattern and have discussed in \S\ref{sec:background_and_motivation}.
    IPCP is the latest composite prefetcher that attempts to capture multiple access patterns including Constant Stride, Complex Stride and Global Stream.
    The last two prefetchers are based on delta-correlation~\cite{Primer}.
    Berti is the latest delta-based prefetcher and can serve the highest L1D accuracy among all evaluated methods to the best of our knowledge. It takes into account the fetch latency to serve a timely and accurate prefetching. We evaluate enhanced Berti, namely vBerti, which works in virtual address and can generate cross-page prefetches.
    The original vBerti allows for prefetching across up to 128 virtual pages (considering both forward and backward directions). However, based on our experiments, this configuration often results in significant performance degradation in multi-core environments {due to high fetch latencies, which prompts vBerti to select larger but less accurate deltas to keep timeliness}.
    Thus, we restrict it to eight virtual pages (four per direction). This adjustment prevents the loss of prefetching opportunities while maintaining its high accuracy. 
    Besides, we also evaluate the widely-used commercial prefetcher IP-stride~\cite{IP-stride}.

    \begin{table}[tb]\renewcommand{\arraystretch}{1.2}
        \centering
        \caption{Configuration and storage overhead of evaluated proposals}
        \label{tab:tab4}
        {
            \begin{tabularx}{3.4in}{m{0.45in} m{2.05in} m{0.4in}}
                \toprule
                SMS     & 2KB region, 64-entry FT/AT, \underline{16k-entry PHT}, \underline{32-entry PB}, \underline{fast access} & 116.6KB \\
                \midrule
                Bingo   & 2KB region, 64-entry FT/AT, \underline{16k-entry PHT}, \underline{32-entry PB}, \underline{fast access} & 138.6KB \\
                \midrule
                DSPatch & 2KB region, 64-entry PageBuffer, 256-entry SPT, \underline{32-entry PB}                                 & 4.25KB  \\
                \midrule
                PMP     & 4KB region, 64-entry FT/AT, 64-entry OPT, 32-entry PPT, 32-entry PB, MaxConf 32, L1/L2 Thresh 0.5/0.15  & 5.0KB   \\
                \midrule
                IPCP    & 64-entry IP table, 128-entry CSPT, 8-entry RST, 32-entry RR                                             & 0.7KB   \\
                \midrule
                SPP-PPF & Same configuration as in~\cite{SPP-PPF}                                                                 & 39.3KB  \\
                \midrule
                vBerti  & Virtual address, eight-page prefetch range                                                              & 2.55KB  \\
                \bottomrule
            \end{tabularx}
        }
        \vspace{-10pt}
    \end{table}

    For fairness, unless specified, all evaluated prefetchers are placed alone at \underline{L1D}. Each prefetcher is implemented based on its source code provided by the authors. The parameters and the storage overhead of all evaluated prefetchers is listed in Table~\ref{tab:tab4}, where the enhanced parts of SMS, Bingo and DSPatch are highlighted. For SMS and Bingo, we use 16k-entry PHTs for their optimal performance. However, an access to such huge PHT takes more than ten CPU cycles. This severely restricts their performance when deployed at L1D. Thus, we assume such an access can be completed within a single CPU cycle.
    For DSPatch, we replace its out-of-dated Prefetch Buffer (PB). The PBs used in the four spatial-pattern-based methods as well as in Gaze have been fine-tuned and are uniform.

    \subsubsection{Main Metrics}
    Below we list the metrics of interest with a brief introduction.

    {\bf Speedup.} The speedup is calculated as the ratio of IPC (instructions per cycle) achieved when with prefetching to that when without prefetching.

    {\bf Accuracy.} We use \textit{overall accuracy} rather than L1D accuracy. This is because some prefetchers choose to put blocks with lower-confidence into lower cache levels (e.g., L2C). These blocks \textit{cannot} be detected by L1D and thus will not be included in L1D accuracy by definition. However, they still play a crucial role in hiding long off-chip latencies and should therefore be considered.
    All the evaluated methods do not put data into LLC. Assume that the number of useful and useless prefetched blocks at L1D are $n_a$ and $n_b$, respectively; the number of useful and useless prefetches at L2C are $m_a$ and $m_b$. Then the overall accuracy is $\frac{n_a + m_a}{n_a + n_b + m_a + m_b}$.

    {\bf Coverage.}
    The coverage refers to the ratio of demands that are successfully covered by prefetching.
    As off-chip misses mainly contribute to performance degradation, we evaluate the LLC coverage.

    {\bf Timeliness.} Since Gaze waits for an additional access, the timeliness may be impacted, i.e., there are more late prefetches. When a CPU access hits on an outstanding prefetch request, we say this prefetch is late~\cite{prefetch_throttling_3}. We measure the fraction of the late prefetches. A small fraction (e.g., 10\%) does not harm the performance, as they still save most of CPU cycles~\cite{prefetch_throttling_3}. 

    \subsection{Results}
    \label{sec:evaluation_results}
        
        \subsubsection{Single-core Performance}
\begin{figure*}[tb]
            \centering
            \includegraphics[scale=1]{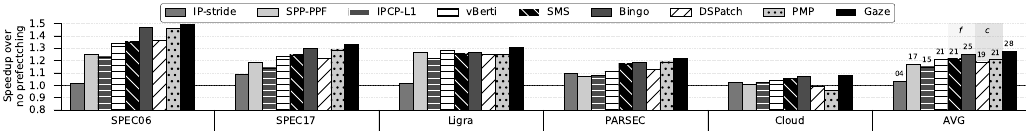}
            \caption{Speedup achieved by the {nine} evaluated prefetchers on each benchmark suite and the overall average (AVG). Group \texttt{f} and \texttt{c} means spatial-pattern-based prefethcers based on fine-grained events and coarse-grained events, respectively.}
            \label{fig:fig6}
            \vspace{-10pt}
        \end{figure*}

\begin{figure}[tb]
            \centering
            \includegraphics[scale=1]{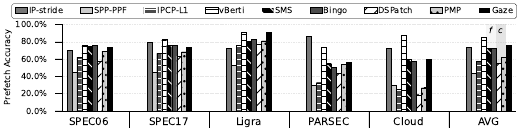}
            \caption{Prefetch accuracy of evaluated prefetchers on each benchmark suite and the overall average (AVG).}
            \label{fig:fig7}
            \vspace{-10pt}
        \end{figure}

\begin{figure}[htb]
            \centering
            \includegraphics[scale=1]{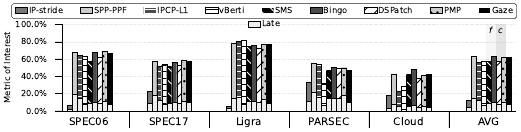}
            \caption{LLC miss coverage and timeliness of evaluated prefetchers on each benchmark suite and the overall average (AVG). Each bar represents the covered portion, where the lower part stands for the late but useful.}
            \label{fig:fig8}
            \vspace{-10pt}
        \end{figure}

        We measure the IPC improvement, accuracy, and coverage of all evaluated prefetchers on \texttt{SPEC06}, \texttt{SPEC17}, \texttt{Ligra}, \texttt{PARSEC} and \texttt{CloudSuite} benchmark suites. The results for each suite in single-core simulation are presented in the Figure~\ref{fig:fig6},~\ref{fig:fig7} and~\ref{fig:fig8}, respectively.
        SMS, Bingo and DSPatch have been enhanced as described in \S\ref{sec:methodology}, enabling them to achieve their optimal performance when working at the L1D.
        
        {\bf Speedup.} Figure~\ref{fig:fig6} presents the speedup achieved by all evaluated prefetchers in the single-core system, where group \texttt{f} and \texttt{c} means previous fine-grained and coarse-grained spatial-pattern-based prefethcers, respectively. As depicted in the figure, Gaze achieves the highest performance improvement across \underline{all} workload suites, with a notable {27.7\%} performance improvement over a no-prefetching baseline on average. Gaze outperforms the second-best prefetcher, Bingo, by {1.9\%}, and significantly surpasses the other two state-of-the-art prefetchers, PMP and vBerti, by {5.7\%} and {5.4\%} respectively. Furthermore, Gaze exhibits superior performance compared to DSPatch and SMS by {7.5\%} and {5.4\%}, and outperforms IPCP-L1, and SPP-PPF by {11.2\%} and {9.1\%}.
        
        Only the two fine-grained prefetchers, Bingo and SMS, along with our novel Gaze, have demonstrated excellent performance (over 5\%) on scale-out \texttt{CloudSuite}, which is consists of server workloads. The other two coarse-grained spatial-pattern-based prefetchers, PMP and DSPatch, even result in performance degradation.\footnote{This is in line with the \texttt{CloudSuite} results mentioned in~\cite{PMP}.} Thus, their simple coarse-grained pattern characterization schemes are inefficient in this scenario.
        
        The \texttt{Ligra} benchmark suite contains various graph algorithms, which involve numerous irregular accesses, as the neighbors of a certain vertex are typically distributed discretely. However, almost all prefetchers successfully achieve outstanding performance improvements on it. We believe that this is because \texttt{Ligra} is well-optimized for memory accessing.
        
        \textbf{Accuracy.} Figure~\ref{fig:fig7} present the detailed prefetch accuracy of all evaluated prefetchers.
        As shown in Figure~\ref{fig:fig7}, Gaze shows the second-highest prefetch accuracy among all prefetchers, surpassing SPP-PPF, IPCP-L1, SMS, Bingo, DSPatch and PMP by 73.6\%, 32.0\%, 4.7\%, 3.6\%, 37.6\%, and 22.5\%, respectively. Surprisingly, compared to Bingo, where substantial hardware overhead is needed to fully utilize longer events \texttt{PC+Address} and \texttt{PC+Offset}, Gaze still outperforms it. This superiority can be attributed to its efficient pattern characterization ability.
        It should be noted that the overall accuracy of vBerti is slightly lower than its L1D accuracy reported in~\cite{Berti}.\footnote{For vBerti, our evaluation also shows an L1D accuracy close to 90\%, which is in line with~\cite{Berti}.} This is because vBerti choose to prefetch low-confidence blocks into L2C, which the L1D cannot detect. Thus, the L1D accuracy does not include them.
        
        Compared to the highly accurate vBerti, the accuracy of Gaze is only 4.2\% lower when excluding \texttt{CloudSuite}. In contrast to other spatial prefetchers, vBerti and IP-stride exhibit excellent accuracy on \texttt{CloudSuite}, indicating that the access patterns of cloud workloads are strongly code-correlated. However, they only cover a small fraction of misses on it. For DSPatch, its second-order enhancement can offer accuracy-oriented prefetching when with only limited available bandwidth. However, the simple bitwise AND operation is insufficient to provide significant gains. Both PMP and Gaze are mainly based on \texttt{Offset}, yet Gaze achieves significantly higher accuracy compared to PMP. Although PMP expands the number of used historical patterns to 32, this does not fundamentally address the accuracy degradation caused by the flawed pattern characterization. In contrast, Gaze achieves a substantial accuracy improvement.
        
        {\bf Coverage.} As shown in Figure~\ref{fig:fig8}, Gaze achieves a moderate coverage, exceeding the highly accurate vBerti by 6.6\%, and remaining at the same level as Bingo and PMP. Gaze achieves nearly no coverage penalty in spite of its strict matching mechanism, which avoids issuing prefetches when the event only partially matches. Gaze still exceeds DSPatch, which is capable of providing coverage-oriented prefetching, by 7.5\%.
        This shows that the eliminated prefetches of Gaze are almost entirely inaccurate. On top of that, vBerti fails to cover enough misses in \texttt{CloudSuite}, where its speedup falls below that of Gaze, Bingo and SMS.
        
        {\bf Timeliness.}
        As shown in Figure~\ref{fig:fig8}, across all traces, Gaze achieves the second-best timeliness {(IP-stride is discarded)}. Gaze only incurs {0.5} percentage points more late prefetches than vBerti ({12.3\%} vs. {11.8\%}). However, vBerti achieves this by extending all L1D lines, all MSHRs, and all entries in Prefetch Queue (PQ) to compute fetch latencies. But this only brings marginal benefits since a small fraction does not harm performance~\cite{prefetch_throttling_3}. Compared to the other four spatial-pattern-based prefetchers, the additional enhancement of Gaze does not degrade the timeliness. We attribute this to its high accuracy, which allows accurate prefetches to be issued more quickly, minimizing the blocking effect of useless ones. IPCP-L1 and SPP-PPF demonstrate poor timeliness, which is unacceptable for high-performance prefetching.
        
        \subsubsection{{Performance Analysis}}
        \label{subsubsec:performance_analysis}
        
        The performance improvement of Gaze can be attributed to two main factors: the efficient pattern characterization, which further aligns with temporal property of memory accessing, and the spatial-streaming-towarded fine-grained prefetching control.
        
        \begin{figure}[b]
            \vspace{-10pt}
            \centering
            \includegraphics[scale=1]{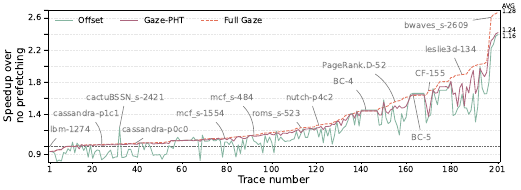}
            \caption{{Effect of our pattern characterization scheme. We present the speedup brought by the two characterization settings and the full Gaze across all traces.}}
            \label{fig:fig9}
        \end{figure}
        
        Figure~\ref{fig:fig9} shows the effect of our new characterization scheme. It plots the performance improvements brought by naively using the trigger offset (Offset), and by leveraging information from the first two accesses (Gaze-PHT), respectively. 
        The speedup of full Gaze, i.e., the synergy of pattern characterization and optimized streaming prefetching, is also plotted. The averages are marked at the end of each curve.

        The workloads on the left side of Figure~\ref{fig:fig9} exhibit minimal spatial regularities. 
        Consequently, simply relying on the trigger offset is ineffective in characterizing their access behaviors, leading to significant pattern misuse.
        In contrast, Gaze-PHT successfully captures some prefetch opportunities.
        On the right side, data prefetching brings noticeable performance improvements, as these workloads typically exhibit various spatial regularities, including spatial streaming. 
        The gap between Offset and Gaze-PHT indicates that, in these regular workloads, waiting for accurate characterization does not miss prefetching opportunities but enhances gains. However, we can also see the underutilization of streaming behavior.

        \begin{figure}[tb]
            \centering
            \includegraphics[scale=1]{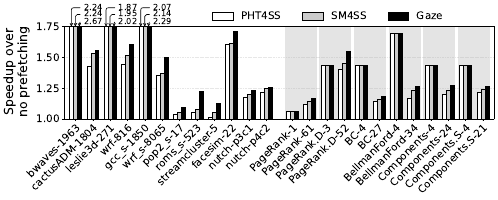} 
            \caption{Effect of the streaming module. We evaluate the speedup brought by the two streaming settings and the full Gaze on several representative workloads.}
            \label{fig:fig10}
            \vspace{-10pt}
        \end{figure}
        
        \begin{figure*}[b]
            \centering
            \includegraphics[scale=1]{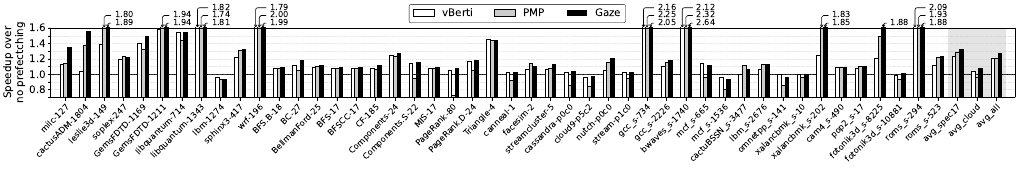}
            \caption{{Detailed speedup achieved by vBerti, PMP and Gaze on representative traces and the averages on different categories.}}
            \label{fig:fig11}
        \end{figure*}
        
        Figure~\ref{fig:fig10} shows the effect of the streaming-towarded enhancement on several representative workloads. We evaluate two streaming settings, one that naively uses PHT for spatial streaming (PHT4SS) and the other that employs our dedicated streaming module (SM4SS), i.e., DPCT and DC.
        To highlight the effectiveness on the streaming behavior, both only operate in streaming regions (i.e., whose initial two accessed blocks are the block 0 and block 1). The speedup of full Gaze is also plotted. 
        
        The left side shows the results on some individual workload traces while the right part shows the results on several \texttt{Ligra} workloads. Each plotted \texttt{Ligra} workload contains two representative traces: one from the initial phase (with smaller suffix numbers) and the other from the computing phase (with larger suffix numbers). During the initial phase, the program is likely preparing data and therefore mainly exhibits streaming behavior, which can be easily captured. Consequently, the performance of PHT4SS, SM4SS, and Gaze are nearly the same. In contrast, during the computing phase, as analyzed in \S\ref{subsec:Enhancement towards Spatial Streaming}, various access patterns co-exist. Thus, naively relying on PHT may misuse the dense pattern. Therefore, the finer-grained streaming module performs much better for streaming behavior.

        \subsubsection{Comparative Study}
        
        We present the performance improvements achieved by Gaze and the two latest spatial prefetchers, PMP and vBerti, across several representative workload traces in Figure~\ref{fig:fig11}.
        
        Compared to PMP and Gaze, vBerti shows relatively lower performance gains in workloads where spatial streaming exists, such as \texttt{leslie3d}, \texttt{bwaves\_s}, and \texttt{xalancbmk\_s}.
        We find that a key factor is the occurrence of redundant prefetches, where prefetch requests are issued for blocks already resident in the L1D.
        Spatial-pattern-based prefetching effectively avoids them by awakening prefetching in newly activated regions. In contrast, vBerti cannot naturally perform such checks, leading to significant redundant prefetches when data are repeatedly accessed. These redundant requests occupy the PQ, only to be discarded when the L1D explicitly checks them. Although this only slightly increases queuing latency, the issuance of normal prefetches is severely hindered, as the PQ may become saturated more frequently. 
        To demonstrate its impact, we implement an oracle version of vBerti (Oracle\_vBerti) that can determine whether a requested block is already present in the L1D before issuing a prefetch. If the block is present, Oracle\_vBerti simply drops the prefetch request. On \texttt{bwaves\_s}, the achieved speedup increases from 2.12 to 2.65. However, this optimization is not a panacea. We also observe a small performance degradation (-4.2\%) on \texttt{GemsFDTD}. Our further analysis reveals that, in many cases, the elimination of redundant prefetches allows more low-confidence prefetch requests to be added into the PQ, leading to a decrease in prefetch accuracy. Overall, the lack of mechanism to handle redundant prefetches hinders its performance to some extent.

        PMP performs well on workloads with simple access patterns but causes significant performance declines on some workloads with complex access behaviors, such as \texttt{canneal}, \texttt{PageRank}, and \texttt{cassandra}. This can be attributed to that using the trigger offset alone can not accurately characterize access behavior. 
        Meanwhile, PMP lacks control over the resulting high aggressiveness.
        
        As shown, Gaze efficiently leverages both simple and complex access patterns. Moreover, the performance degradation caused by Gaze on some irregular workloads (e.g., \texttt{cloud9} and \texttt{mcf\_s}) is slight.
        We observe performance decreases of up to 6.9\% across all traces, where PMP and vBerti result in maximum performance declines of 27.3\% and 8.5\%, respectively. 
        Thus, the performance variability of Gaze is reasonable.
        Table~\ref{tab:tab5} summarizes the comparison of Gaze, vBerti, PMP and Bingo.
        
        \begin{table}[tb]  
            \centering  
            \caption{Comparison of Prefetchers} 
            \label{tab:tab5}  
            \begin{tabular}{|c|c|c|c|}  
                \hline  
                    \textbf{Prefetcher} & 
                        \makecell[c]{
                            \specialrule{0em}{0.5pt}{0.5pt}
                            \textbf{Hardware}\\\textbf{Cost}\\
                            \specialrule{0em}{0.5pt}{0.5pt}
                        } 
                    & 
                    \makecell[c]{
                        \specialrule{0em}{0.5pt}{0.5pt}
                        \textbf{Simple Pattern}\\(e.g., streaming)\\
                        \specialrule{0em}{0.5pt}{0.5pt}
                    } 
                    &
                    \makecell[c]{
                        \specialrule{0em}{0.5pt}{0.5pt}
                        \textbf{Complex Pattern}\\(e.g., cloud)\\
                        \specialrule{0em}{0.5pt}{0.5pt}
                    } \\ 
                \hline  \hline
                    \makecell[c]{
                        \specialrule{0em}{0.5pt}{0.5pt}
                        \textbf{Gaze} \\
                        \specialrule{0em}{0.5pt}{0.5pt}
                    }  & \ding{52} & \ding{52} & \ding{52} \\ 
                \hline  
                    \textbf{vBerti}    & 
                    \makecell[c]{
                        \specialrule{0em}{0.5pt}{0.5pt}
                        \ding{52} \\ \\
                        \specialrule{0em}{0.5pt}{0.5pt}
                    }
                     & 
                        \makecell[c]{
                            \specialrule{0em}{0.5pt}{0.5pt}
                            \halfcheck \\ Redun. req.\\
                            \specialrule{0em}{0.5pt}{0.5pt}
                        } &
                        \makecell[c]{
                            \specialrule{0em}{0.5pt}{0.5pt}
                            \halfcheck \\ Low cov.\\
                            \specialrule{0em}{0.5pt}{0.5pt}
                        } 
                        \\ 
                \hline  
                    \textbf{PMP}       & 
                    \makecell[c]{
                        \specialrule{0em}{0.5pt}{0.5pt}
                        \ding{52} \\ \\
                        \specialrule{0em}{0.5pt}{0.5pt}
                    } & 
                    \makecell[c]{
                        \specialrule{0em}{0.5pt}{0.5pt}
                        \ding{52} \\ \\
                        \specialrule{0em}{0.5pt}{0.5pt}
                    } & 
                        \makecell[c]{
                            \specialrule{0em}{0.5pt}{0.5pt}
                            \ding{56} \\ Low acc.\\
                            \specialrule{0em}{0.5pt}{0.5pt}
                        } 
                        \\ \hline  
                    \makecell[c]{
                        \specialrule{0em}{0.5pt}{0.5pt}
                        \textbf{Bingo} \\
                        \specialrule{0em}{0.5pt}{0.5pt}
                    }       & \ding{56} & \ding{52} & \ding{52} \\ 
                \hline  
            \end{tabular}  
        \end{table}  
        
        In \texttt{lbm}, all three main prefetchers achieve high accuracy (over 90\%) and cover more than one-third of LLC misses, yet still result in performance degradations. This can be contributed to increased queuing delays for CPU demands at the DRAM controller.
        In fact, some prefetches are issued too early. Thus, a feasible solution could be to prioritize CPU demands during bandwidth bottlenecks.
        Meanwhile, Gaze also experiences slight performance declines due to insufficient accuracy in a few traces, such as \texttt{xalancbmk\_s}.
        Since active regions are tracked by AT, Gaze can observe whether the prefetch pattern matches actual accesses. Based on this, Gaze can set confidence for each pair of initial offsets, and can reject low-confidence prefetch patterns. We have implemented similar control for streaming behavior in Gaze. The extensions for other patterns are leaved for future work.
        
        \subsubsection{GAP and QMM Performance}

        We further present the performance of the three major prefetchers on a subset of \texttt{GAP} workloads~\cite{GAP} and Qualcomm (\texttt{QMM}) industry workloads. \texttt{GAP} is a framework for graph analytics, similar to \texttt{Ligra}. 
        \texttt{QMM} contains numerous anonymized industry traces released by Qualcomm in the first Championship Value Prediction (CVP-1)~\cite{CVP1}. The \texttt{QMM}  traces we use are converted by~\cite{qmm_converter}.
        
\begin{figure}[tb]
            \centering
            \includegraphics[scale=1]{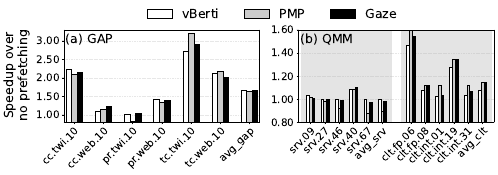} 
            \caption{Speedup achieved for (a) GAP and (b) QMM benchmark suites by vBerti, PMP and Gaze. QMM is further divided into server (left part) and client (right part) workloads.}
            \label{fig:fig12}
            \vspace{-5pt}
        \end{figure}

        Our evaluation on \texttt{GAP} contains three graph algorithms (\texttt{cc}, \texttt{pr}, and \texttt{tc}) on two real-world datasets (\texttt{twitter} and \texttt{web-sk-2005}). As depicted in Figure~\ref{fig:fig12}a, similar to \texttt{Ligra}, \texttt{GAP} is sensitive to data misses. Gaze and vBerti both achieve outstanding improvements. However, PMP shows severe performance decline on the irregular trace.
        On average, Gaze surpasses vBerti and PMP by 1.3\% and 2.7\%, respectively.
        
        The \texttt{QMM} workloads can be furthre divided into two categories: server workloads (plotted on the left part of Figure~\ref{fig:fig12}b), which typically include large code footprints, and client workloads (plotted on the right part), which primarily consist of memory-intensive computing tasks. As shown, spatial prefetching achieves substantial performance improvements on client workloads but nearly on improvements on server workloads. This is because the primary bottleneck for \texttt{QMM} server workloads is instruction misses instead of data misses~\cite{Morrigan}. They have relatively higher L1I MPKI values and lower LLC MPKI values, compared to \texttt{CloudSuite}, which means that they are much less sensitive to data misses.
        In \texttt{QMM} server workloads, although Gaze successfully covers nearly one-third of the LLC misses, a slight decline is still observed (-1.6\% on average). Highly accurate vBerti also fails to provide effective gains (+0.4\%), while PMP brings significant performance drop (-10.2\%).

        \subsubsection{Multi-level Prefetching}
        
\begin{figure}[b]
            \vspace{-5pt}
            \centering
            \includegraphics[scale=1]{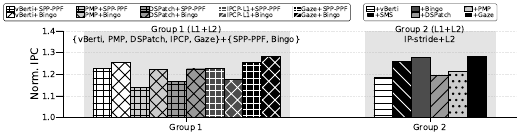} 
            \caption{Speedup achieved in multi-level prefetching. Group 1 contains the eight combinations of recent prefetchers. Group 2 shows their performance when combined with IP-stride.}
            \label{fig:fig13}
        \end{figure}

        We evaluate eight combinations of several state-of-the-art prefetchers (Group 1). We also present their performance when combined with the current commercial L1D prefetcher IP-stride (Group 2).
        Figure~\ref{fig:fig13} shows the speedup achieved by these combinations.

        The combination Gaze+Bingo achieves the best performance improvement, with an increase of {0.34\%} in IPC compared to only using Gaze at L1D. The other nine combinations still fall short of using Gaze alone.
        Bingo, PMP, DSPatch and Gaze all are bit-pattern-based. Consequently, they can act as filters when working together with Bingo, resulting in marginal performance improvements. In fact, performance degradations are even observed in several combinations due to the additional aggressiveness introduced by L2 prefetching. Delta-correlated vBerti shows room for performance enhancement when integrated with spatial-pattern-based Bingo.
        Meanwhile, we do not see significant improvements when combining them with SPP-PPF.
        
        When combined with the commercial L1D prefetcher IP-stride, Gaze still achieves the best performance. The vBerti is highly dependent on the completeness of access sequences, which can be a limitation when operating at the L2C.
        Likewise, when Gaze is deployed at L2C alongside an aggressive L1D prefetcher capable of generating cross-page prefetching, the true initial accesses may be difficult to obtain. In this case, similar to IPCP~\cite{IPCP}, transmitting metadata through MSHRs is a feasible solution. However, as we do not see any considerable benefit from multi-level prefetching, using Gaze alone at L1D is a better choice.
        
        \begin{figure}[tb]
            \centering
            \includegraphics[scale=1]{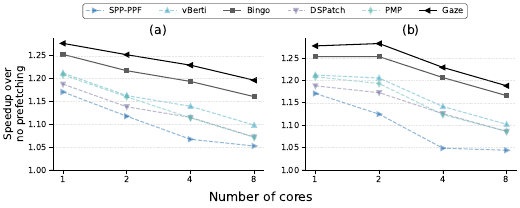} 
            \caption{Multi-core performance achieved in (a) homogeneous and (b) heterogeneous mixes.}
            \label{fig:fig14}
            \vspace{-10pt}
        \end{figure}

        \subsubsection{Multi-core Performance}

        Figure~\ref{fig:fig14} depicts the performance improvements in homogeneous and heterogeneous multi-core simulations. Thanks to the higher accuracy, the performance degradation of Gaze is more gradual. Gaze consistently achieves the highest performance across 1-8 cores. 
        The two aggressive coarse-grained prefetchers, DSPatch and PMP, exhibit significantly performance degradation. The vBerti shows a relatively smoother degradation compared to PMP, DSPatch and SPP-PPF.
        Thus, among all prefetchers, Gaze can be applied in a wider range of systems, delivering excellent performance both in scenarios with abundant memory bandwidth resources and higher bandwidth contention.

        We first focus on homogeneous results (Figure~\ref{fig:fig14}a) since the heterogeneous mixes involve interactions among different workloads. 
        As the number of cores increases, bandwidth contention intensifies. 
        Thanks to the higher accuracy, the degradation trends of Gaze, Bingo, and vBerti are more gradual. However, for PMP, the performance degradation becomes particularly severe when the number of cores increases to four or more. This is due to its inefficient characterization and excessive aggressiveness.
        In eight-core homogeneous mixes, Gaze outperforms Bingo, PMP, and vBerti by 3.1\%, 11.7\%, and 9.0\%, respectively.

        \begin{table}[h]\renewcommand{\arraystretch}{1.5}
            \centering
            \caption{{Selected four-core mixes}}
            \label{tab:tab6}
            {
                \begin{tabularx}{3.1in}{m{0.2in} m{2.85in}}
                    \hline
                    \textbf{Mix1} & wrf-1254, Triangle-1, lbm\_s-2676, Triangle-6 \\
                    \hline
                    \textbf{Mix2} & GemsFDTD-1211, PageRank-19, BFS.B-5, BFS-5                                       \\
                    \hline
                    \textbf{Mix3}  & bwaves\_s-2609, BFSCC-1, wrf\_s-8065, astar-359                                                           \\
                    \hline
                    \textbf{Mix4}  & PageRank.D-24, bwaves-1963, PageRank-61, facesim-22                                                         \\
                    \hline
                    \textbf{Mix5}  & cass.-p0c0, cass.-p0c1, cass.-p0c2, cass.-p0c3                                                  \\
                    \hline
                \end{tabularx}
            }
        \end{table}

\begin{figure}[tb]
            \centering
            \includegraphics[scale=1]{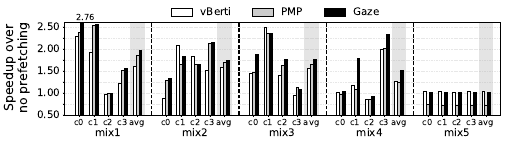} 
            \caption{{Four-core performance achieved in different mixes.}}
            \label{fig:fig15}
            \vspace{-10pt}
        \end{figure}

        For heterogeneous simulations, we plot the prefetching performance on several representative four-core mixes (listed Table~\ref{tab:tab6}). As illustrated in Figure~\ref{fig:fig15}, Gaze shows advantages in both per-core performance and the overall average.
        It is noteworthy that, within a mix, the effectiveness of prefetching varies significantly across different cores. This can be attributed to the fact that heterogeneous workloads compete for shared resources  
        in different manners. As a result, weaker workloads may struggle to secure sufficient resources, which can be further exacerbated in the presence of prefetching.
        Some approaches attempt to introduce global control at the LLC to address this issue~\cite{prefetch_throttling_1,prefetch_throttling_2,prefetch_throttling_3}.
        In eight-core heterogeneous mixes, Gaze outperforms  PMP and vBerti by 9.4\% and 7.8\%, respectively.
        
        \begin{figure}[b]
            \vspace{-5pt}
            \centering
            \includegraphics[scale=1]{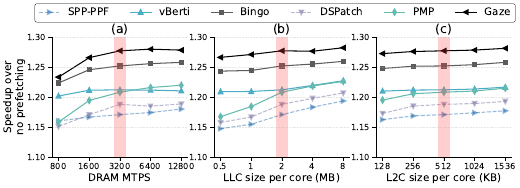} 
            \caption{Sensitivity to (a) DRAM bandwidth, (b) LLC size and (c) L2C size. The baseline configuration is marked in red.}
            \label{fig:fig16}
        \end{figure}
        
        \subsubsection{Sensitivity to System Configurations} 
        
        As shown in Figure~\ref{fig:fig16}, we evaluate the performance gains when with varying off-chip bandwidth, LLC sizes {and L2C sizes}. Gaze performs well in both high- and low-bandwidth environments, and can adapt nicely to various LLC{/L2C} sizes.
        The vBerti performs well under limited resources but cannot scale effectively when available resources are adequate. The over-aggressiveness of PMP leads to a extreme sharp decline when bandwidth or cache size decreases. 
        Meanwhile, the results indicate that the lightweight Gaze can effectively scale to high-bandwidth environments, such as DDR5~\cite{ddr}, and adapt to new cache technologies such as 3D V-Cache~\cite{3d_cache}.

        \subsubsection{Sensitivity to Gaze Configurations}
        
\begin{figure}[tb]
            \centering
            \includegraphics[scale=1]{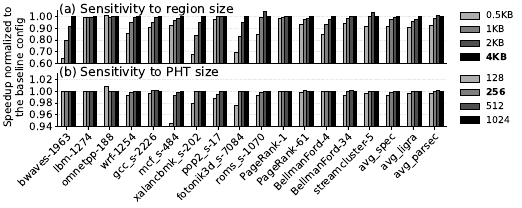} 
            \caption{{Performance of Gaze with different configurations.}}
            \label{fig:fig17}
        \end{figure}

        Figure~\ref{fig:fig17}a illustrates the effect of different region sizes. The results are normalized to the baseline configuration (i.e., 4KB region). Using larger regions almost consistently yields performance improvements, particularly on workloads that are easy to follow. In contrast, using smaller regions frequently misses prefetch opportunities. On average, employing 0.5KB, 1KB, and 2KB regions results in performance degradation of 9.1\%, 4.4\%, and 1.6\%, respectively.

        Figure~\ref{fig:fig17}b presents the effect of PHT sizes. A large table can accommodate more experiences, therefore capturing more prefetching opportunities. As shown, for some workloads, the 128-entry PHT suffers from its limited capacity while 256-entry is sufficient to capture recent spatial patterns. Further increasing the capacity several-fold only brings marginal performance gains. Perversely, for workload \texttt{omnetpp}, 128-entry PHT performs the best. This is because Gaze cannot utilize its complex patterns effectively, and smaller tables reduce the number of mispredictions.
        On average, compared to the baseline configuration, using a 128-entry PHT leads to a 0.6\% performance decrease, while increasing the table size to 512 and 1024 results in only 0.1\% performance gains.

        \begin{figure}[tb]
            \centering
            \includegraphics[scale=1]{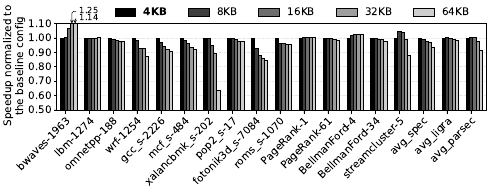}
            \caption{Performance of vGaze with larger region sizes.}
            \label{fig:fig18}
            \vspace{-10pt}
        \end{figure}
        
        Modern operating systems (OSes) commonly use huge pages~\cite{large_page} to reduce the address translation overhead by increasing the TLB hit rate and lowering the TLB miss penalty~\cite{huge_page_1,huge_page_2}, which makes prefetching at larger region sizes a feasible direction.
        Gaze can work at larger region sizes without specific architectural support since virtual addresses are visible at the L1D. As shown in Figure~\ref{fig:fig18}, we evaluate the performance of virtual Gaze (vGaze) with different large region size settings, normalized to the baseline configuration (4KB vGaze).
    
        As illustrated, among the plotted traces, only \texttt{bwaves} can benefit noticeably from larger region sizes. As we have discussed in \S\ref{subsubsec:performance_analysis}, this speedup can be attributed to its streaming access. However, for most workloads, using region sizes larger than 4KB only leads to performance degradation. This suggests that most spatial patterns better align with the 4KB region size. 
        Consequently, naively using larger regions is ineffective.
        
        We further conduct the evaluation of accuracy and coverage. The results show that, when using small regions (less than 4KB), the accuracy remains relatively stable, but the coverage experiences a sharp decline. This reveals that smaller regions cannot effectively utilize spatial locality. In contrast, when enlarging the region size from 4KB to up to 32KB, coverage marginally grows, but the accuracy declines relatively faster. Our evaluation indicates that a region size of 4KB is suitable for Gaze.
        
        Meanwhile, a recent work has shown that smartly switching the granularity based on the hint of OS page size from TLB can be effective~\cite{page_size_aware_prefetching}. Since Gaze can work in virtual address, such support is unnecessary for cross-4KB-boundary prefetching. However, multiple PHTs with different settings are required to support region size switching. It is orthogonal to the design of Gaze, and we leave it as future work.

\section{Other Related Work}
\label{sec:relatedwork}

\textbf{Prefetching} has been extensively studied over decades.
\textit{Delta-correlated spatial prefetchers} including BOP~\cite{BOP}, SPP~\cite{SPP}, and Berti~\cite{Berti} try to predict future deltas. BOP records the scores of every delta and selects the one with the highest score. Berti works in a per-PC view and further takes into account the fetch latency to serve a timely and accurate prefetching. SPP and IPCP-CPLX~\cite{IPCP} use signature-based methods to recognize complex strides.
\textit{Temporal prefetchers}~\cite{Domino,Triage,MISB,STeMS,Markov,ISB,STMS,GHB} try to extract temporal similarity of access patterns. They are proficient at recognizing irregular access patterns such as pointer-based access. However, they suffer from the huge storage overhead since the demand sequences are stored.
Some prefetchers apply \textit{machine learning} in hardware prefetching~\cite{Pythia,ML_Pref_1,ML_Pref_2,ML_Pref_3}, where Pythia~\cite{Pythia}, a reinforcement learning-based hardware prefetcher, achieves both high performance and acceptable hardware complexity.
\textit{Pre-computation}~\cite{pre_exec_1,pre_exec_2,pre_exec_3,pre_exec_4,pre_exec_5,pre_exec_6,pre_exec_7,pre_exec_8} issues prefetch requests through pre-executing program code. They can generate highly-accurate prefetches without the need for access patterns. However, pre-executing dramatically increases the hardware complexity.
\textit{Cache level prediction}~\cite{TLP,Hermes,CLP_1,CLP_2,CLP_3,CLP_4} hides the on-chip cache hierarchy latency by predicting which level a load might hit and fetch the data directly. Hermes~\cite{Hermes} uses a perceptron-based mechanism to predict which demands might go off-chip.

{\bf Spatial footprint prediction} is also leveraged in several DRAM cache techniques~\cite{FootprintCache_1, FootprintCache_2, FootprintCache_3, FootprintCache_4} to harness the advantages of both cacheline- and page-based designs. They manage data at a coarse granularity to avoid infeasible large tag space, while fetching only the blocks that will be used during the residency of the page to overcome the over-fetching problem.

\section{Conclusion}
\label{sec:conclusion}

In this paper, we highlight that the underlying limitation of spatial-pattern-based prefetching is its pattern characterization scheme.
To this end, we propose Gaze, a lightweight and efficient hardware prefetcher that Gaze skillfully applies the temporal correlation of the first several spatial accesses to characterize patterns, which makes it capable of predicting more accurately with low hardware cost.
Concurrently, Gaze introduces a dedicated two-stage approach to mitigate the over-prefetching problem that often occurs in spatial streaming.
In our evaluation, Gaze adapts well to both low-bandwidth and high-bandwidth environments and exhibits excellent performance in multi-core simulations. Gaze outperforms PMP and vBerti by 5.7\% and 5.4\% at single-core and by 11.4\% and 8.8\% at eight-core.

\section*{Acknowledgements}
\label{sec:ACKNOWLEDGMENTS}
We thank all the anonymous reviewers, as well as our shepherd, for their valuable comments. 
This work was supported in part by the National Key R\&D Program of China (Grant No. 2023YFB4502902).

%
%
%
%
%

\appendix
\section{Artifact Appendix}

\subsection{Abstract}
We implement Gaze using ChampSim simulator~\cite{ChampSim}.
In this artifact, we provide all the necessary information to reproduce the main experiments presented in the paper. We describe how to prepare the required datasets, set up the system, and run the artifact. This artifact includes the source code of the ChampSim simulator used in our experiments and the scripts to build all binaries, run the experiments, and draw figures. 

\subsection{Artifact check-list and meta-information}

{\small
\begin{itemize}
  \item {\bf Program:} ChampSim
  \item {\bf Compilation:} G++ 11.4.0
  \item {\bf Data set:} Traces from SPEC06~\cite{SPEC06}, SPEC17~\cite{SPEC17}, Ligra~\cite{Ligra}, PARSEC~\cite{PARSEC}, CloudSuite~\cite{CloudSuite}, QMM~\cite{CVP1}, and GAP~\cite{GAP} benchmark suites.
  \item {\bf Hardware:} No specific hardware requirements.
  \item {\bf Output:} 15 PDF figures.
  \item {\bf How much disk space required?} $\sim$100GB
  \item {\bf How much time is needed to prepare workflow?} $\sim$2 hours, including downloading traces.
  \item {\bf How much time is needed to complete experiments?} $\sim$7-10 days when using a compute cluster with 384 cores: key single-core results (1 day, fig. 1, 6-8, 11), multi-core results (4-5 days, fig. 14-15), supplementary single-core results (2-3 days, fig. 4, 9, 10, 12-13, 16-18).
  \item {\bf Publicly available?} Yes
  \item {\bf Code licenses (if publicly available)?} GPL-3.0
  \item {\bf Archived (provide DOI)?} \href{https://doi.org/10.5281/zenodo.14252372}{10.5281/zenodo.14252220}
\end{itemize}
}

\subsection{Description}

\subsubsection{How to access}
The source code can be obtained from GitHub: \href{https://github.com/SJTU-Storage-Lab/Gaze-Spatial-Prefetcher.git}{https://github.com/SJTU-Storage-Lab/Gaze-Spatial-Prefetcher.git}

\subsubsection{Hardware dependencies} 
No specific hardware requirements. Gaze can be run on any system with a general-purpose CPU.

\subsubsection{Software dependencies}
Gaze is tested on Ubuntu 18.04. The following dependencies are recommended:

\begin{itemize}
    \item Ubuntu 18.04.6 LTS
    \item Linux Kernel 5.4.0-150-generic
    \item G++ 11.4.0
    \item Python 3.8.18
    \begin{itemize}
        \item matplotlib 3.7.2
        \item numpy 1.24.3
        \item pandas 2.0.3
        \item scipy 1.10.1
    \end{itemize}
\end{itemize}

\subsubsection{Data sets}
The traces used in our experiments are primarily from championships or previous work. Below is the detailed information about obtaining these traces. They should be manually downloaded. 

\begin{itemize}
    \item \textbf{SPEC06 and SPEC17 traces.} Available from DPC-3~\cite{DPC3} (\url{https://dpc3.compas.cs.stonybrook.edu/}).
    \item \textbf{PARSEC and Ligra traces.} Originally from Pythia~\cite{Pythia} (not available now), we have archived a copy of our used traces.
    \item \textbf{CloudSuite traces.} Available from CRC-2~\cite{CRC2} (\url{https://crc2.ece.tamu.edu/}).
    \item \textbf{GAP traces} (only used in supplementary experiments). Archived.
    \item \textbf{QMM traces} (only used in supplementary experiments). Archived.
\end{itemize}

All traces should be placed directly in the trace directory \texttt{\$GAZE\_HOME/traces}.

Run the following command to check if the required traces are ready.
\begin{verbatim}  
    $ cd $GAZE_HOME/scripts   
    $ python trace_check.py
\end{verbatim} 

\subsection{Installation}

\vspace{5pt}
1. Clone Gaze from GitHub repository
\begin{verbatim}  
    $ git clone   
\end{verbatim}

Gaze is implemented in ChampSim, thus, ChampSim needs to be successfully built. The following are the build steps from \texttt{ChampSim/README.md}. For more detailed information, please refer to it.

\vspace{5pt}
2. Download ChampSim's dependencies
\begin{verbatim}  
    $ cd ChampSim
    $ git submodule update --init
    $ vcpkg/bootstrap-vcpkg.sh
    $ vcpkg/vcpkg install  
\end{verbatim} 

Then, we can compile all prefetchers.

\vspace{5pt}
3. Build all binaries
\begin{verbatim}  
    $ cd ../scripts/make
    $ python make_all.py
\end{verbatim} 

\subsection{Experiment workflow}
This section lists instructions to execute experiments. 
The running and plotting scripts are located in \texttt{scripts/run} and \texttt{scripts/draw}, respectively.
The simulation will generate raw log and json files that will be used for figure plotting.
Some eight-core heterogeneous simulations take more than four days to complete, therefore, please allocate dedicated CPU resources for multi-core heterogeneous simulations.
Each script launches all corresponding simulations at once, which may occupy excessive system resources. It can be modified to run only a subset of the simulations.
After manually confirming that the relevant simulations have completed, the plotting script can be executed to generate the figure. 

\vspace{5pt}
1. Main single-core simulations (fig. 1, 6-8, 11)
\begin{verbatim}  
    $ cd $GAZE_HOME/scripts/run
    $ python run_single_core_main.py
    $ # wait for simulation
    $ cd ../draw
    $ python fig1.py # fig/fig1.pdf
\end{verbatim} 

\vspace{5pt}
2. Multi-core homogeneous simulations (fig. 14a)
\begin{verbatim}  
    $ python run_multi_core_homo.py
\end{verbatim} 

\vspace{5pt}
3. Multi-core heterogeneous simulations (fig. 14b, 15)
\begin{verbatim}  
    $ python run_multi_core_hete.py
\end{verbatim} 

4. Supplementary simulations (fig. 4, 9, 10, 12-13, 16-18)
\begin{verbatim}  
    $ python the_running_script.py 
\end{verbatim} 

\subsection{Evaluation and expected results}

The simulation does not directly output results. Instead, it writes the raw data to a json file, which is saved in \texttt{GAZE\_HOME/json} directory. Meanwhile, runtime logs are stored in \texttt{GAZE\_HOME/log}.
The plotting scripts convert the raw data into result data and generate the figures included in the paper.
Therefore, the generated figures should be consistent with those presented in the paper.

\subsection{Notes}
Before generating a figure, please ensure all corresponding simulations have completed. This can be manually verified through process status or the associated logs (in \texttt{log}).
This is because, in a few multi-core simulations, ChampSim actively terminate due to invalid simulation states, which can result in invalid raw data. Thus, the plotting code is designed to handle invalid log and json files. However, not all invalid files are generated from active termination; some are invalid because the simulations remain unfinished.

\bibliographystyle{IEEEtranS}
\bibliography{refs}

\end{document}